\documentclass[aps,pre,twocolumn,amsmath,amssymb,showpacs,floatfix,reprint]{revtex4-1}
\pdfoutput=1
\usepackage{overpic}
\usepackage[T1]{fontenc}
\usepackage{amsmath}
\usepackage{amssymb}
\usepackage{lmodern}
\usepackage{units}
\usepackage{color}
\usepackage{graphicx}
\usepackage{booktabs}

\newcommand*{\ellk}{\ell_{\rm K}}
\newcommand*{\dhh}{D_{\rm H}}
\newcommand*{\dw}{D_{\rm W}}
\newcommand*{\lcc}{l_{\rm cc}}
\newcommand*{\lch}{l_{\rm ch}}
\newcommand*{\lcw}{l_{\rm cw}}

\newcommand*{\rb}{R_{\rm blob}}
\newcommand*{\fc}{F_{\rm c}}

\begin{document}
\title{Scaling regimes of a semi-flexible polymer in a rectangular channel}
\author{E. Werner, B. Mehlig}
\affiliation{Department of Physics, University of Gothenburg, Sweden}

\begin{abstract}
We derive scaling relations for the extension statistics and the confinement free energy for a semi-flexible polymer confined to a channel with a rectangular cross-section. Our motivation are recent numerical results [Gupta {\em et al.}, JCP {\bf 140} (2014) 214901] indicating that extensional fluctuations are quite different in rectangular channels compared to square channels. Our results are of direct relevance for interpreting current experiments on DNA molecules confined to nano-channels, as many experiments are performed for rectangular channels with large aspect ratios while theoretical and simulation results are usually obtained for square channels.
\end{abstract}
\maketitle

Conformations of semi-flexible polymers confined to narrow channels are presently intensively investigated,
since it has emerged that nano-channel confined DNA molecules offer possibilities both for biological applications and as a model system for experimental polymer physics \cite{tegenfeldt2004,reisner2005,reccius2005,persson2009,werner2012,reisner2012,jones2013,gupta2014,alizadehheidari2015c,reinhart2015}. These studies, and the theoretical and simulations results which have accompanied them, have shown that semi-flexible polymers such as DNA exhibit a much richer spectrum of behaviours under confinement than do flexible polymers \cite{odijk2008,wang2011,tree2012,dai2013,muralidhar2014,muralidhar2014a}.

Since it is convenient to fabricate channels with a fixed height but varying widths, 
many experimental studies \cite{tegenfeldt2004,reccius2005,reisner2005,persson2009,kim2011,utko2011,werner2012,frykholm2014,gupta2014} of confined DNA are performed in rectangular channels 
(width $\dw$, height $D_{\rm H}$) with aspect ratios far from unity, $\dw\!\gg\!\dhh$. However, most simulation and theoretical studies \cite{wang2011,tree2012,werner2012,werner2013,dai2013,muralidhar2014,muralidhar2014a,dai2014b,werner2014} are restricted to channels with square cross-sections, $\dw\!=\!\dhh$. A common procedure is to analyse the experimental results in terms of the \lq effective channel size\rq{} $\sqrt{\dw \dhh}$, simply disregarding the influence of the aspect ratio.

\begin{figure}[t]
\includegraphics[width=7cm]{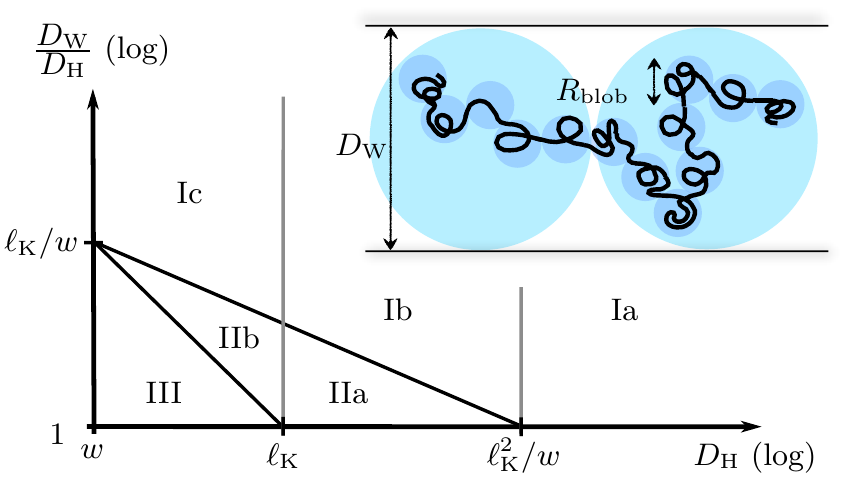}
\caption{\label{fig:phaseDiagram} Phase diagram of different scaling
regimes for the extension statistics of a semi-flexible polymer confined
to a channel with a rectangular cross-section. The scaling in regime I
is given by Eqs. (\ref{eqn:extension_deGennes},\ref{eqn:variance_deGennes}). The scaling in regime II is given by Eqs. (\ref{eqn:extension_extendedDeGennes},\ref{eqn:variance_extendedDeGennes}). The vertical grey lines distinguish different scaling regimes of the confinement free energy, see text. Results for regime III have been derived elsewhere \cite{odijk2008,burkhardt2010,chen2013,muralidhar2014a} and are discussed in the SM.
Inset: An illustration of the hierarchy of blobs analysed in regime I. 
In regime Ia, the smaller blobs are spherical blobs of size $\rb = \dhh$. In regime Ib-c they are cylindrical, of height $\dhh$ and width $\rb > \dhh$.}
\end{figure}

Recently Gupta {\em et al.}~\cite{gupta2014} have shown that the variance of the extension of the DNA molecule does depend on the aspect ratio, but a theoretical explanation for this intriguing result is lacking. This motivated us to analyse how the aspect ratio influences the extension statistics of confined semi-flexible polymers. At
first sight this appears to be a difficult problem, because it is governed
by a large number of length scales: the contour length $L$ of the polymer,
its effective width $w$,
its Kuhn length $\ellk$ \cite{grosberg1994}, its global persistence length $g$ \cite{odijk2008}, 
the typical contour-length separation between intra-chain collisions $\lcc$,
the typical contour-length separation between collisions with the floor and the ceiling of the channel $\lch$,
the typical contour-length separation between collisions with the vertical walls of the channel $\lcw$, and the channel width $\dw$ and height $\dhh$.
There are many different confinement regimes corresponding to different orderings of these length scales, 
potentially resulting in a very complicated phase diagram (Fig.~\ref{fig:phaseDiagram}). 
Little is known about this phase diagram for
rectangular channels, except in the limit of very strong confinement \cite{odijk2008,burkhardt2010,chen2013}.
To interpret a given experiment it is necessary to determine which regime in the phase diagram it corresponds to, and what the resulting scaling relations
for the extension statistics are.

In this paper we summarise the results of our analysis, based on the mean-field theory for
the extension of an unconfined semi-flexible polymer \cite{flory1953}.
It is well known how this theory must be adapted to describe the extension of
semi-flexible polymers confined to square channels with $\dw\!=\!\dhh\!\equiv\!D$ \cite{brochard1977}: 
for a wide channel and a long polymer, the polymer globule is divided
into a series of smaller spherical blobs of size $D$. One assumes that mean-field scaling holds 
for each blob and concludes that the extension of the confined polymer scales as $D^{-2/3}$.
We generalise this analysis to rectangular channels. To this end it is necessary to consider a hierarchy 
of blobs (inset of Fig.~\ref{fig:phaseDiagram}). For the special case of a flexible chain
this approach was  used by Turban to compute its extension in a rectangular channel \cite{turban1984}. 
We emphasise that this is a much simpler problem since a flexible chain exhibits only a single confinement regime, as opposed
to the semi-flexible polymer, as pointed out above.

An important result of our analysis is that the scalings can be simply summarised, despite the fact that the phase diagram Fig.~\ref{fig:phaseDiagram} exhibits many different regimes. First, we find that the average extension is approximately independent of 
channel aspect ratio, provided that at least one of the channel dimensions is significantly larger than the Kuhn length of the polymer. This is an important finding because it implies that it is approximately correct to analyse the average extension
of nano-confined DNA molecules in terms of the effective channel size $\sqrt{\dw\dhh}$.
Second, we find that the extension variance depends strongly on both channel dimensions separately, it would
be incorrect to analyse it in terms of the effective channel size. We find that the variance
increases rapidly with aspect ratio if far from unity. Our theoretical results for the variance explain the findings of Ref.~\cite{gupta2014}, they also make it clear that square channels are much preferred for applications
where extensional fluctuations are required to be as small as possible. Third, we compute
the free energy of confinement, it is approximately determined by the smallest confining dimension. 
These scaling predictions were derived under the mean-field approximation of Flory \cite{flory1953}, but in certain parameter regimes (regimes IIa-b in Fig.~\ref{fig:phaseDiagram}), the results are supported by an asymptotically exact theory that was developed for square channels \cite{werner2014}, but can be generalised to rectangular channels.

We have summarised these results in a table in the supplemental material (SM) \cite{SM}, also including results for very strong confinement
that were derived by other authors.
For other observables (e.g. dynamics, statistics of circular polymers, probability of knot formation), the scaling properties remain to be determined, but the distinctions between the regimes we derive here must also apply to them.

{\em Method.} 
We assume that the polymer is described by the self-avoiding worm-like chain model \cite{grosberg1994}
with Kuhn length $\ellk$ and effective width $w < \ellk$.
The centre-line of the polymer is confined to a rectangular channel of height $\dhh$ and width $\dw\ge \dhh$. 
We analyse the conformational fluctuations of the confined 
polymer in terms of a hierarchy of blobs assuming that self-avoiding random walks obeying Flory's
mean-field theory \cite{flory1953} in two and three spatial dimensions. To summarise this theory 
assume that the polymer consists of $N$ independent segments of length $\ell$ and excluded volume $v$.
In $d=2,3$ dimensions the mean-field result for the average of the extension $R$ of the polymer globule is  \cite{flory1953}:
\begin{align}
\label{eqn:FloryExtensionGeneral}
R \approx \left(N^3 \ell^2 v\right)^\frac{1}{d + 2}\,.
\end{align}
Here the symbol $\approx$ indicates a scaling relation, ignoring factors of order unity.
The variance of the extension is determined by self similarity. The self-avoiding polymer in two or three
dimensions has only one 
macroscopic length scale ($R$) \cite{grosberg1994}. Thus, the standard deviation of the extension 
must scale as the average extension \cite{degennes1976}:
\begin{align}
\label{eqn:FloryDeviationGeneral}
\sigma_R \approx R \approx \left(N^3 \ell^2 v\right)^\frac{1}{d + 2}.
\end{align}

As we show below, the analysis
of the extension statistics of the confined polymer must proceed by different steps, depending on the relation between the contour length scales $\ellk ,\, g ,\, \lcc ,\, \lch ,\, \lcw$, defined in the introduction. The global persistence length $g$ only differs appreciably from the Kuhn length $\ellk$ at very strong confinement (regime III in Fig.~\ref{fig:phaseDiagram}).
Theories for this regime
have been derived elsewhere 
\cite{odijk2008,burkhardt2010,chen2013,muralidhar2014a}, and
are briefly discussed in the SM \cite{SM}.

{\em Extension statistics.} 
Consider the separation of scales
\begin{align}
\label{eqn:separationOfLengthScales_rectangularDeGennes}
\ellk \ll \lcc \ll \lch \ll \lcw \ll L\,.
\end{align}
How do the average extension and its standard deviation depend upon $\dw$ and $\dhh$?
Since $\lcc \ll \lch$, the polymer exhibits three-dimensional Flory scaling before its 
first collision with the channel walls. The first collision with the ceiling or the floor must occur when a section of contour length $\lch$ has formed a spherical Flory blob of diameter $\dhh$. Applying Eq.~(\ref{eqn:FloryExtensionGeneral}) with $N= \lch/\ellk$ and $v\approx\ellk^2 w$ \cite{onsager1949} yields
\begin{align}
\label{eqn:lch_deGennes}
\dhh \approx \left(\lch^3  v/\ellk\right)^\frac{1}{5} \Leftrightarrow 
\lch  \approx [\dhh^5/(\ellk w)]^{1/3}\,.
\end{align}
The resulting blobs perform a two-dimensional self-avoiding walk until they have formed a circular 
\lq superblob\rq{} of diameter $\dw$ (we follow the terminology of Ref.~\cite{turban1984}). 
This two-dimensional random walk is illustrated in the inset of Fig.~\ref{fig:phaseDiagram}. The number of 
small spherical blobs constituting one superblob can be estimated from Eq.~(\ref{eqn:FloryExtensionGeneral}) with $d=2$, and $\ell = \dhh$ since each small 
spherical blob constitutes an independent segment of the walk. We assume
that the small blobs do not overlap,
thus the two-dimensional excluded volume of the random walk of small blobs
approximately equals the area of a circle of diameter $\dhh$, in other words $v\approx \dhh^2$ in this context.
Eq.~(\ref{eqn:FloryExtensionGeneral}) results in
\begin{align}
\dw \approx \left(N_{\rm blobs}^3 \dhh^4\right)^\frac{1}{4} \Leftrightarrow N_{\rm blobs} \approx (\dw/\dhh)^{4/3}\,.
\end{align}
The contour length stored within a superblob equals $\lcw$,
\begin{align}
\label{eqn:lcw_deGennes}
\lcw \approx N_{\rm blobs} \lch \approx [\dw^4 \dhh /(\ellk w)]^{1/3}.
\end{align}
There are $L/\lcw$ superblobs that line up along the channel 
[Fig.~\ref{fig:phaseDiagram}(inset)]. Each superblob has average diameter $\dw$ with fluctuations of the same order. The average extension and its fluctuations are therefore given by:
\begin{align}
\label{eqn:extension_deGennes}
R &\approx (L/\lcw) \dw \approx L \left[{\ellk w}/({\dhh \dw})\right]^{1/3}, \\
\sigma_R^2 &\approx (L/\lcw) \dw^2 \approx L \left({\ellk w \dw^2}/{\dhh}\right)^{1/3}\,.\label{eqn:variance_deGennes}
\end{align}
Eq.~(\ref{eqn:extension_deGennes}) was recently derived in Ref.~\cite{benkova2015a}, but without specifying under which conditions the 
derivation is valid. 
We now 
answer this question.
The inequality $\lcc \ll \lch$ requires that ideal scaling within a blob of size $\dhh$ results in a large number of intra-chain collisions within the blob
\begin{align}
(\dhh/\ellk)^4 v/\dhh^3  \approx \dhh w /\ellk^2 \gg 1 \Leftrightarrow \dhh \gg \ellk^2/w.
\end{align}
This corresponds to
regime Ia in Fig.~\ref{fig:phaseDiagram}.

The inequality $\lcw \gg \lch$ is satisfied if $\dw \gg \dhh$. However, the scaling relations 
(\ref{eqn:extension_deGennes},\ref{eqn:variance_deGennes}) reproduce exactly the well-known relations for square channels 
as the limit $\dw \to \dhh$ is taken, commonly referred to
as \lq de Gennes scaling\rq{} \cite{brochard1977} .
Thus de Gennes scaling for square channels is simply a special case of the more general scaling relations derived above for rectangular channels.
Yet attempting to generalise from square to rectangular channels by simply replacing $D$ by the geometrical average $(\dhh \dw)^{1/2}$ gives the wrong prediction for the variance of the extension, as Eq.~(\ref{eqn:variance_deGennes}) shows.

Now consider a different ordering of length scales:
\begin{align}
\label{eqn:separationOfLengthScales_intermediateDeGennes}
\ellk \ll \lch \ll \lcc \ll \lcw \ll L\,.
\end{align}
This corresponds to
$\dhh \ll \ellk^2/w$ while $\dw$ remains large, 
regime Ib in Fig.~\ref{fig:phaseDiagram}. Are
the scaling relations (\ref{eqn:extension_deGennes}) and (\ref{eqn:variance_deGennes}) modified in this regime?
Since $\lch \ll \lcc$, the polymer must exhibit ideal scaling even after the first collision with the ceiling or the floor of the channel. This scaling persists approximately until the first intra-chain collision, which 
occurs after a contour length $\lcc$. This length scale is estimated by assuming that such a section of the polymer will form a cylindrical blob with height $\dhh$ and diameter $\rb\approx\sqrt{\lcc \ellk}$, and by setting the expected number of collisions within the blob to unity:
\begin{align}
\label{eqn:lcc_intermediateDeGennes}
\frac{(\lcc/\ellk)^2 \ellk^2 w}{\rb^2 \dhh} = \frac{\lcc w}{\ellk \dhh} = 1 \Leftrightarrow \lcc = \frac{\ellk \dhh}{ w}.
\end{align}
These blobs perform a two-dimensional self-avoiding walk until the first collision with the side walls
[Fig.~\ref{fig:phaseDiagram}(inset)]. Flory scaling for the $N = \lcw/\lcc$ blobs of size $\ell\approx\rb\approx\sqrt{\lcc \ellk}$ and excluded area $v\approx \rb^2\approx \lcc \ellk$ yields
\begin{align}
\lcw \approx[\dw^4 \dhh /(\ellk w)]^{1/3}\,.
\end{align}
This is the same as in regime Ia, Eq.~(\ref{eqn:lcw_deGennes}).
From this point, the derivation of the extension statistics follows  that of regime Ia.
We infer that  the scalings Eqs.~(\ref{eqn:extension_deGennes},\ref{eqn:variance_deGennes}) hold also in regime Ib.

Further decreasing $\dhh$ below $\ellk$ one enters a different regime (labelled Ic in Fig.~\ref{fig:phaseDiagram}). It corresponds to this ordering of length scales:
\begin{align}
\label{eqn:separationOfLengthScales_quasi2DDeGennes}
\lch \ll \ellk \ll \lcc \ll \lcw \ll L\,.
\end{align}
In this case the polymer runs almost completely parallel to the floor of the channel but can otherwise rotate freely. While the expression for the excluded volume of a Kuhn length segment must change because of confinement, the scaling $v\approx \ellk^2w$ still holds \cite{onsager1949}. The same analysis as in regime Ib can be carried out here, leading to identical scaling predictions for the extension statistics, Eqs.~(\ref{eqn:extension_deGennes},\ref{eqn:variance_deGennes}). The prefactors will differ between the regimes, however.

Since the steps in the above derivation are different between regime Ia and regimes Ib-c, it is at first glance surprising that the scaling predictions for the extension are identical for these regimes. But note that the scaling analysis
is formulated in terms of blobs that obey Flory's mean-field scaling. As long as mean-field theory is used throughout, the final prediction for the extension must be independent of the way in which the blobs are arranged. 
Since in mean-field theory the repulsive effect of self-avoidance is directly determined by the number of monomers within the volume spanned by the polymer, this also explains why the scaling of the extension is a function of the cross-section only, independent of the aspect ratio of the channel. 

What about the variance of the extension? Combining mean-field theory with the universality of self-avoiding random walks \cite{degennes1976} shows that the contour length contained in one superblob [Fig.~\ref{fig:phaseDiagram}(inset)] is identical for these regimes, and that each blob experiences size fluctuations of order $\dw$. The variance of the extension is given by summing the variances of each blob, yielding $\sigma_R^2 \approx N_{\rm superblobs} \dw^2 \approx L (\ellk w\dw^2/\dhh)^{1/3}$. Thus the variance increases as the aspect ratio increases. This demonstrates that rectangular channels are in fact quite different from square ones in this regime.

If the excluded volume of the polymer is so small that the polymer experiences multiple collisions with side walls, floor, and ceiling between each intra-chain collision, then we obtain different scaling relations. Consider the following ordering of length scales
\begin{align}
\label{eqn:separationOfLengthScales_extendedDeGennes}
\ellk \ll \lch \le \lcw \ll \lcc \ll L\,.
\end{align}
It corresponds to $\ellk \ll \dhh \le \dw \ll (\dhh \ellk^2/w)^{1/2}$ (regime IIa in Fig.~\ref{fig:phaseDiagram}).
Under these conditions the polymer obeys ideal scaling until a blob forms that fills the channel cross-section, and is further elongated along the channel direction, until it reaches an extension $\rb \approx \sqrt{\lcc \ellk}$. 
The length scale $\lcc$  is estimated in a similar way as Eq.~(\ref{eqn:lcc_intermediateDeGennes}) is obtained:
\begin{align}
\frac{(\lcc/\ellk)^2 \ellk^2 w}{\rb \dhh \dw} \approx 1 \Leftrightarrow \lcc \approx \left(\frac{\ellk \dhh^2\dw^2}{ w^2}\right)^{1/3}\!\!\!\!.
\end{align}
The polymer arranges itself into $L/\lcc$ blobs  of size $\rb$:
\begin{align}
\label{eqn:extension_extendedDeGennes}
R &\approx (L/\lcc) \rb 
\approx L \left[{\ellk w}/({\dhh \dw})\right]^{1/3}\!\!\!\!, \\
\sigma_R^2 &\approx (L/\lcc) \rb^2 \approx L \ellk\,.\label{eqn:variance_extendedDeGennes}
\end{align}
We see that the scaling of the average extension agrees with (\ref{eqn:extension_deGennes}). The reason is that both equations were derived assuming mean-field statistics within each blob, and as noted above, the ordering of the blobs does not influence the prediction for the average extension. However, note that the scaling of the standard deviation, 
Eq.~(\ref{eqn:variance_extendedDeGennes}) differs from Eq.~(\ref{eqn:variance_deGennes}). We see that $\sigma_R$
does not depend upon either $\dw, \dhh$, or $w$. 

For the special case of square channels, regime IIa has been studied under the name \lq extended de Gennes regime\rq{} \cite{wang2011,dai2013,dai2014b,werner2014}.
Fig.~\ref{fig:phaseDiagram} shows that the limits of this regime are more restrictive for rectangular  than for square channels: even if a square channel with side length either $\dhh$ 
or $\dw$ is in regime IIa, the rectangular channel with side lengths $\dhh$ and $\dw$ might not be.

Finally consider the ordering of length scales 
\begin{align}
\label{eqn:separationOfLengthScales_quasi2DExtendedDeGennes}
\lch \ll \ellk \ll \lcw \ll \lcc \ll L\,,
\end{align}
corresponding to $\dhh\ll\ellk \ll \dw \ll (\dhh \ellk^2/w)^{1/2}$. This regime is denoted as IIb in Fig.~\ref{fig:phaseDiagram}. The steps needed to derive
the scaling relations for the extension are identical to those summarised above, and again lead to Eqs.~(\ref{eqn:extension_extendedDeGennes},\ref{eqn:variance_extendedDeGennes}), albeit with different prefactors than for regime IIa. The scaling for the extension in this regime was previously derived by Odijk \cite{odijk2008}. 

{\em Comparison with results of computer simulations.} Our results give a qualitative explanation for the surprising observation of Gupta {\em et al.} \cite{gupta2014}, who found in simulations that while the average extension of a confined polymer was relatively insensitive to the aspect ratio of the confining channel, the variance increased with aspect ratio (see Fig.~5 in Ref.~\cite{gupta2014}). They performed simulations with a fixed channel height $\dhh = 100\,{\rm nm}$ and channel widths in the range $\dw = 100 - 1000 \,{\rm nm}$, and compared these against simulations of square channels with matching effective channel size. For their simulations, they used a polymer with $\ellk= 137.4\, {\rm nm}$ and $w= 18.7\, {\rm nm}$. Thus, while the square channels are all approximately in regime IIa, the channels with fixed height cross over into 
regime Ib as the aspect ratio increases significantly above unity (Fig.~\ref{fig:phaseDiagram}). While our prediction for mean extension shows the same scaling in these regimes, the variance is independent of $\dhh$ and $\dw$ in regime IIa but increases as $\dw^{2/3}$ in regime Ib, qualitatively explaining the results of Gupta {\em et al.}. That the agreement is only qualitative is not surprising, considering that the conditions for regime Ib are only marginally satisfied, and that the contour length of $\lambda$-DNA which they replicate is not quite long enough to enter the asymptotic regime where $r\propto L$.

{\em Accuracy of mean-field theory.}
The results that we derived for regimes I and II are based on Flory's mean-field theory, which is thought to be correct in two dimensions but is known to be only approximate in three dimensions \cite{li1995}. Using the scaling $R\propto L^{0.588}$ in three dimensions would lead to modified scaling predictions for regime Ia (but not for the other regimes): $R\propto \dhh^{-0.37} \dw^{-1/3}$, $\sigma^2\propto \dhh^{-0.37} \dw^{2/3}$, $\fc\propto \dhh^{-1.70}$.

For square channels in regime IIa, we have recently shown both that the scalings of mean-field theory are exact in this regime, and that there are rigorous bounds for the prefactors \cite{werner2014}. These results were derived by mapping the statistics of the extended de Gennes regime to a one-dimensional model known as the weakly self-avoiding random walk. Since the same mapping can be performed for rectangular channels, these exact results are valid throughout regime IIa. The scalings of Eqs.~(\ref{eqn:extension_extendedDeGennes},\ref{eqn:variance_extendedDeGennes}) can thus be rigorously proven. The rigorous bounds for the prefactors are included in the table in the SM \cite{SM}.
As in regime IIa, it is in principle possible to map the statistics of regime IIb to a solved one-dimensional model. Computing the exact parameters of the mapping would require performing an integral over the monomer density profile in the Odijk regime, yet even without it the existence of the mapping shows that also in this regime, the scalings of the extension statistics are exact.

{\em Free energies.}
Apart from the statistics of the extension, the free energy of confinement $\fc$ is of experimental relevance, as it determines the force that must be applied to introduce a polymer into a channel. This free energy can be estimated by $k T$ times the number of collisions with the walls, or
\begin{align}
\fc / (k T) \approx L/\lch + L/\lcw \approx L/\lch.
 \label{eqn:freeEnergy_general}
\end{align}
For regime Ia, we obtain a scaling prediction for 
$\fc$ by inserting 
Eq.~(\ref{eqn:lch_deGennes}) into Eq.~(\ref{eqn:freeEnergy_general}), yielding $\fc\approx L (\ellk w/\dhh^5)^{1/3}$. For all other regimes, the free energy of confinement is the same as for ideal polymers, $\fc\approx L\ellk/\dhh^2$ in regimes Ib and IIa \cite{casassa1967}, and $\fc\approx L(\ellk \dhh^2)^{-1/3}$ in regimes Ic and IIb \cite{burkhardt2010}.
More exact predictions for $\fc$ are included in the table in the SM \cite{SM}. 

{\em Conclusions.} Recent experiments on DNA in rectangular nano-channels are performed at high aspect ratio, yet most analytical and simulation results pertain to square channels. These analytical results are often applied to rectangular channels by matching the cross-sectional area of the rectangular channel to that of the square one. Our theory shows that this matching allows one to predict the average DNA extension under fairly general assumptions. We also show that this procedure fails to correctly predict the scalings of other observables, such as the variance. Our theory explains recent numerical results for the extension variance in rectangular channels \cite{gupta2014}, and shows that square channels are most useful for biological applications where it is beneficial that the extension variance is small. We expect that the results summarised here can be generalised to other observables, such as the statistics of circular DNA, knot formation, and DNA dynamics.

{\em Acknowledgements.}
Financial support from Vetenskapsr\aa{}det and from the G\"oran Gustafsson Foundation for Research in Natural Sciences and Medicine is gratefully acknowledged.

\end{document}


\title{{\rm Supplemental material for}\\ Scaling regimes of a semi-flexible polymer in a rectangular channel}
\author{E. Werner, B. Mehlig}
\affiliation{Department of Physics, University of Gothenburg, Sweden}
\maketitle
We summarise our predictions for the  scalings for the statistics of the extension, and the free energy of confinement in Table \ref{tab:1}. 
Where numerical prefactors are known, they have been included in the table.
In this table, we also include previous results for the scalings, at very strong confinement (regime III of Fig.~1), and for a polymer confined to a slit (the limit $\dw\to \infty$). These results are briefly discussed below. 

\subsubsection{Odijk regime}
If both $\dhh \ll \ellp$ and $\dw \ll \ellp$, then it is impossible for the polymer to turn around completely. Instead, the polymer must stay almost parallel to the channel axis, undulating slightly from side to side. In this regime the statistics of the extension and the free energy of confinement 
have been studied in detail both theoretically and numerically \cite{burkhardt2010,chen2013}.

\subsubsection{Backfolded Odijk regime}
Odijk \cite{odijk2008} has predicted the existence of a regime intermediate between the Odijk regime and the extended de Gennes regime, for very slender semi-flexible polymers. For square channels, this regime was studied
by Muralidhar {\em et al.} \cite{muralidhar2014a}, who gave it the name ``backfolded Odijk regime''. In this regime the size of the channel is such that backfolds are possible but rare. Odijk defines the \textit{global persistence length} $g$ as the orientational correlation length of the corresponding ideal polymer. The backfolded Odijk regime requires that $\ellp \ll g\ll \lcc$. This condition can only be satisfied for very thin polymers \cite{muralidhar2014a}.
Assuming that a polymer section that is free of backfolds 
follows the statistics of the Odijk regime, one can predict scaling relations for the extension \textit{in terms of the global persistence length} $g$. However, no theory for how $g$ itself depends on the channel size exists for relevant channel sizes \cite{muralidhar2014a}. Note also that the statistics of the Odijk regime are not derived under the assumption that the chain does not backfold, but under the stricter assumption that it is almost completely parallel with the channel axis. Since such strong alignment prohibits backfolding, the agreement with Odijk statistics is only approximate in this regime.

The predictions for the backfolded Odijk regime in the square channel \cite{odijk2008,muralidhar2014a} are straightforward to generalise to rectangular channels. The results are shown in Table \ref{tab:1}. The prediction for the scaling of the extension was derived by Odijk \cite{odijk2008}. 

\subsubsection{Slit regimes}
For completeness, we include in Table \ref{tab:1} the scaling regimes for infinite channel aspect ratio, 
i.e. for the polymer confined in a slit. The results for these regimes are quoted from recent studies by Dai {\em et al.} \cite{dai2012}, 
Taloni {\em et al.} \cite{taloni2013}, and Tree {\em et al.} \cite{tree2014}.
For polymers that are too short to exhibit the asymptotic relation $R\propto L$ (see main text),
these scalings also describe the statistics in regimes Ia-c in Fig.~1 in the main text.

\begin{turnpage}

\begin{table*}    
\mbox{}\hspace*{-1cm}\begin{tabular}{llllllll}
\toprule
&& \multicolumn{3}{c}{Conditions} & \multicolumn{3}{c}{Statistics} \\
\cmidrule(lr){3-5}
\cmidrule(lr){6-8}
Regime & Name    & $\dhh$ & $\dw$ \footnote{We assume throughout that $\dw\ge\dhh$.} & $L$ & $R/L$ & $\sigma_R^2/L$ & $\fc/L$\\
\midrule
Ia & de Gennes      				& \small $\gg\ellk^2/w$ & --- & $\gg \Big(\frac{\dw^4 \dhh} {\ellk w}\Big)^{\frac{1}{3}}$ & $\approx \Big(\frac{\ellk w}{\dhh \dw}\Big)^{\frac{1}{3}}$ \cite{benkova2015a} & $\approx \Big(\frac{\ellk w \dw^2}{\dhh}\Big)^{\frac{1}{3}}$ \cite{werner2015}& $\approx \Big(\frac{\ellk w}{\dhh^5}\Big)^{\frac{1}{3}}$ \cite{werner2015} \\
Ib & --- & $\ellk\ll \dhh \ll \frac{\ellk^2}{w}$  & $\dw^2 \gg \dhh \ellk^2/w$  & $\gg \Big(\frac{\dw^4 \dhh} {\ellk w}\Big)^{\frac{1}{3}}$ & $\approx \Big(\frac{\ellk w}{\dhh \dw}\Big)^{\frac{1}{3}}$ \cite{werner2015} & $\approx \Big(\frac{\ellk w \dw^2}{\dhh}\Big)^{\frac{1}{3}}$ \cite{werner2015}& $=\frac{\pi^2}{6} \ellk\dhh^{-2}$ \cite{casassa1967}\\
Ic & --- & $w\le\dhh\ll\ellk$ \footnote{\mbox{Tree et al. \cite{tree2014} have shown in simulations that for polymers confined in slits, the non-crossing regime $\dhh<w$} \mbox{   is approached smoothly as $\dhh\to w$ from above, and we expect the same to be valid in channels.}}  & $\dw^2 \gg \dhh \ellk^2/w$   & $\gg \Big(\frac{\dw^4 \dhh} {\ellk w}\Big)^{\frac{1}{3}}$ & $\approx \Big(\frac{\ellk w}{\dhh \dw}\Big)^{\frac{1}{3}}$ \cite{werner2015}& $\approx \Big(\frac{\ellk w \dw^2}{\dhh}\Big)^{\frac{1}{3}}$ \cite{werner2015}& $= 1.1032(1) \ellp^{-\frac{1}{3}} \dhh^{-\frac{2}{3}}$ \cite{chen2013}\footnote{\mbox{Whereas predictions involving $\ellp$ are specific to the worm-like chain, those expressed in terms of $\ellk$ are valid for general} \mbox{    polymer models, if the effective width $w$ is replaced by $v/\ellk^2$, where $v$ is the excluded volume of a Kuhn length segment.}} \\
IIa & extended de Gennes 			& $\ellk\ll \dhh \ll \frac{\ellk^2}{w}$ & $\dw^2 \ll \dhh \ellk^2/w$ & $\gg\Big(\frac{\dhh^2\dw^2\ellk}{w^2}\Big)^{\frac{1}{3}}$ & $= 0.9338(84)\Big(\frac{\ellk w} {\dhh \dw}\Big)^\frac{1}{3}$ \cite{werner2014} & $=0.13(1) \ellk $ \cite{werner2014} & $=\frac{\pi^2}{6} \ellk(\dhh^{-2} + \dw^{-2})$ \cite{casassa1967} \\
IIb & --- & $w\ll\dhh\ll\ellk$  & $\ellk^2\ll\dw^2 \ll \frac{\dhh \ellk^2}{w}$   & $\gg\Big(\frac{\dhh^2\dw^2\ellk}{w^2}\Big)^{\frac{1}{3}}$  & $\approx \Big(\frac{\ellk w}{\dhh \dw}\Big)^{\frac{1}{3}}$ \cite{odijk2008} & $ = 0.20(2) \ellk $ \cite{werner2014}\footnote{The step variance $\sigma_0^2$ \cite{werner2014} obeys $\sigma_0^2= \ellk^2/2$ in this regime.} & $= 1.1032(1) \ellp^{-\frac{1}{3}} \dhh^{-\frac{2}{3}}$ \cite{chen2013} \\
IIIa & Odijk     					& $\dhh \ll \ellp$ & $\dw \ll \ellp$ & $ \gg \ellp^{\frac{1}{3}}\dw^{\frac{2}{3}}$ & $= 1\! - 0.09137(7) \frac{\dhh^{\frac{2}{3}} + \dw^{\frac{2}{3}} }{\ellp^{\frac{2}{3}}}\! $ \cite{burkhardt2010} & $= 0.00478(10)  \frac{\dhh^2 + \dw^2}{\ellp}\! $ \cite{burkhardt2010} & $= 1.1032(1) \ellp^{-\frac{1}{3}} (\dhh^{-\frac{2}{3}}\! + \dw^{-\frac{2}{3}})\! $ \cite{chen2013} \\
IIIb & backfolded Odijk 			& \multicolumn{2}{c}{$\dhh\le\dw\lesssim \ellp,\; \frac{g w}{\ellp^{\frac{1}{3}}\dw^{\frac{2}{3}}\dhh}\ll 1$ \footnote{\mbox{The global persistence length $g$ is the orientational correlation length of the confined polymer, defined in Refs.~\cite{odijk2008,muralidhar2014a}. }}} & $\gg \Big(\frac{g^{\frac{3}{2}}\ellp\dhh^3\dw^{2}}{w^3}\Big)^\frac{2}{9}$  & $\approx \Big(\frac{g w}{\ellp^{\frac{1}{3}}\dw^{\frac{2}{3}}\dhh}\Big)^{\frac{1}{3}}$ \cite{odijk2008} &  $\approx g$ \cite{muralidhar2014a} & $\approx \ellp^{-\frac{1}{3}} \dhh^{-\frac{2}{3}}$ \cite{muralidhar2014a} \\
\midrule
\multicolumn{2}{l}{Slit regimes} \\
\midrule
Sa & de Gennes      				& $\gg\ellk^2/w$ & ---  & $\gg \Big(\frac{\dhh^5}{\ellk w}\Big)^{\frac{1}{3}}$ & $\approx \Big(\frac{\ellk w}{L\dhh}\Big)^\frac{1}{4}$ \cite{dai2012} & $\approx \Big(\frac{L\ellk w}{\dhh}\Big)^{\frac{1}{2}}$ \cite{taloni2013} & $\approx \Big(\frac{\ellk w}{\dhh^5}\Big)^{\frac{1}{3}}$ \cite{taloni2013} \\
Sb & extended de Gennes			& $\ellk\ll \dhh \ll \frac{\ellk^2}{w}$ & ---  & $\gg \frac{\ellk \dhh}{w}$ & $\approx \Big(\frac{\ellk w}{L\dhh}\Big)^\frac{1}{4}$ \cite{dai2012} & $\approx \Big(\frac{L\ellk w}{\dhh}\Big)^{\frac{1}{2}}$ \cite{taloni2013}& $=\frac{\pi^2}{6} \ellk\dhh^{-2}$ \cite{casassa1967} \\
Sc & Odijk-Flory 				& $w <\dhh \ll \ellk$ & --- & $\gg \frac{\ellk \dhh}{w}$ & $\approx \Big(\frac{\ellk w}{L\dhh}\Big)^\frac{1}{4}$ \cite{odijk2008} & $\approx \Big(\frac{L\ellk w}{\dhh}\Big)^{\frac{1}{2}}$ \cite{taloni2013} & $= 1.1032(1) \ellp^{-\frac{1}{3}} \dhh^{-\frac{2}{3}}$ \cite{chen2013}\\
\bottomrule
\end{tabular}	
\caption{\label{tab:1} Summary of scaling regimes for a semi-flexible polymer confined to a rectangular channel.}

\vspace*{-.7cm}
\end{table*}

\end{turnpage}

\clearpage